\documentclass[10pt,letterpaper]{article}%
\usepackage{opex3}
\usepackage{amsmath}
\usepackage{amsfonts}
\usepackage{amssymb}
\usepackage{graphicx}
\usepackage{relsize}

\usepackage{cite}
\newcommand{\hide}[1]{\relax}
\newcommand{\unit}[1]{\ensuremath{\,\mathrm{#1}}}
\newcommand{\mvec}[1]{\ensuremath{\vec{#1}}}
\newcommand{\Og}{\ensuremath{\Omega}}
\newcommand{\Om}{\ensuremath{\Omega_\mathrm{m}}}
\newcommand{\Omod}{\ensuremath{\Omega_\mathrm{mod}}}
\newcommand{\Gm}{\ensuremath{\Gamma_\mathrm{m}}}
\newcommand{\meff}{m_\text{eff}}
\newcommand{\xzpf}{\ensuremath{x_{\mathrm{zpf}}}}
\newcommand{\dwdx}{\ensuremath{G}}
\newcommand{\vcr}{\ensuremath{g_0}}
\newcommand{\etac}{\ensuremath{\eta_{\mathrm{c}}}}

\begin{document}

\title{Determination of the vacuum optomechanical coupling rate using frequency noise calibration}

\author{M.~L.~Gorodetksy,$^{1,2}$ A.~Schliesser,$^{1,3}$ G.~Anetsberger,$^{3}$ S.~Deleglise,$^{1}$ and T.~J.~Kippenberg$^{1,3}$}
\address{
$^{1}$Ecole Polytechnique F$\acute{e}$d$\acute{e}$rale de Lausanne, EPFL, 1015 Lausanne, Switzerland\\
$^{2}$Faculty of Physics, Moscow State University, Moscow, 119991, Russia\\
$^3$Max-Planck-Institut f{\"u}r Quantenoptik, Hans-Kopfermann-Str. 1, 85748
Garching, Germany 
}

\email{tobias.kippenberg@epfl.ch}

\begin{abstract}
The strength of optomechanical interactions in a cavity optomechanical system can be quantified by a vacuum coupling rate $\vcr$ analogous to cavity quantum electrodynamics. 
This single figure of merit removes the ambiguity in the frequently quoted coupling parameter defining the frequency shift for a given mechanical displacement, and the effective mass of the mechanical mode. 
Here we demonstrate and verify a straightforward experimental technique to derive the vacuum optomechanical coupling rate.
It only requires applying a known frequency modulation of the employed electromagnetic probe field and knowledge of the mechanical oscillator's occupation.
The method is experimentally verified for a micromechanical mode in a toroidal whispering-gallery-resonator and
 a nanomechanical oscillator coupled to a toroidal cavity via its near field.
\end{abstract}
%270.2500   Quantum Optics -Fluctuations, relaxations, and noise
%120.5050   Instrumentation, measurement, and metrology - Phase measurement
%230.4685     Optical devices - Optical microelectromechanical devices

%% activate for two-column option

%\tableofcontents

\bibliographystyle{unsrt}
\bibliography{/Users/aschlies/Documents/Literature/microCavities}
%\bibliography{microCavities}

\section{Introduction}

Optomechanical systems are being explored for a variety of studies, ranging from exploring quantum limits of displacement measurements \cite{Braginsky1975,Caves1981}, ground-state cooling of mechanical oscillators \cite{Marquardt2007,Wilson-Rae2007} to low-noise radiation-pressure driven oscillators \cite{Vahala2008}. 
In this new research field of cavity optomechanics \cite{Kippenberg2008, Marquardt2009}, rapid progress  towards these goals has recently been enabled by the conception and realization of a large variety of nano- and micro-optomechanical systems, which implement parametric coupling of a mechanical oscillator to an optical or microwave resonator (for an overview of recently studied systems, see e.\ g.\ reference \cite{Schliesser2010}).
Thereby, the resonance frequency  $\omega_\mathrm{c}$ of the electromagnetic cavity is changed upon a mechanical displacement $x$ according to $\omega_\mathrm{c}(x)=\omega_\mathrm{c}(0)+\dwdx x$, with the coupling parameter 
\begin{equation}
 \dwdx\equiv \frac{\mathrm{d}\omega_\mathrm{c}}{\mathrm{d} x}.
\end{equation}

The resulting optomechanical interaction can be described by the interaction Hamiltonian \cite{Law1995},
\begin{equation}
\hat{H}_{\mathrm{int}}=
\hbar \dwdx \,\hat xÊ\, \hat n_\mathrm{o}=
\hbar \dwdx \,\xzpf (\hat{a}_{\mathrm{m}}^{\dagger}+\hat{a}_{\mathrm{m}}) \,\, \hat{a}_\mathrm{o}^\dagger \hat{a}_\mathrm{o},
\label{e:hint}
\end{equation}
where $\hat x= \xzpf (\hat{a}_\mathrm{m}^\dagger+\hat{a}_\mathrm{m}) $ is the mechanical displacement operator, $\hat n_\mathrm{o}=\hat{a}_\mathrm{o}^\dagger \hat{a}_\mathrm{o}$ the photon occupation of the optical cavity, and $\hat{a}_\mathrm{o}$, $\hat{a}_\mathrm{o}^\dagger$, $\hat{a}_{\mathrm{m}}$ and $\hat{a}_\mathrm{m}^\dagger$ are the usual ladder operators of the optical (o) and mechanical (m) degrees of freedom.
The  root-mean-square amplitude of the mechanical oscillator's zero-point fluctuations 
\begin{equation}
 \xzpf= \sqrt\frac{\hbar}{2 \meff \Om}
\end{equation}
is determined by its resonance frequency $\Om$ and its effective mass $\meff$ as discussed in more detail below.

The canonical example of such a system would be a Fabry-P\'erot resonator, one mirror of which is mounted as a mass on a spring moving as a whole along the symmetry axis of the cavity.
In this case, the coupling parameter is given by $G=-\omega_\mathrm{c}/L$, where $L$ is the cavity length, and the oscillator's effective mass simply is the moving mirror's mass.
In most real optomechanical systems, however, \emph{internal} mechanical modes constitute the mechanical degree of freedom of interest (for simplicity, our discussion will be restricted to a single mechanical mode only).
In general, the displacement of such a mode has to be described by a three-dimensional vector field $\mvec{u}(\mvec r)$. For every volume element located at $\mvec r$, it yields the displacement $\mvec{u}(\mvec r)$ from its equilibrium position.
In order to retain the simple one-dimensional description of equation (\ref{e:hint}), it is then necessary to map the vector field $\mvec u(\mvec r)$ to a scalar displacement~$x$.

A frequently adapted procedure to address this mapping is to deliberately choose a coupling parameter $G$ adequate to the overall geometry of the system, such as $G=-\omega_\mathrm{c}/L$ for a Fabry-P\'erot-type cavity of length $L$ \cite{Gillespie1995,Gigan2006, Arcizet2008a}, or $G=-\omega_\mathrm{c}/R$ for a whispering-gallery-mode (WGM) resonator of major radius $R$ \cite{Hofer2009,Schliesser2010}.
The equivalent displacement $x$ is then defined by the measured resonance frequency shift $\Delta\omega_\mathrm{c}$ induced by the displacement field $\mvec u(\mvec r)$, i.\ e. $x\equiv\Delta \omega_\mathrm{c}/G$.
Following a different approach, the equivalent displacement $x$ is defined, for example, as the displacement of the structure at a given position, $x\equiv |\mvec u(\mvec r_0)|$ or the maximum displacement of a mode \cite{Ilchenko1994,Regal2008,Jayich2008,Eichenfield2009}. 
Then, the coupling parameter $G$ is calculated via the relation $x=\Delta \omega_\mathrm{c}/G$. In some cases \cite{Braginsky1993,Ilchenko1994,Eichenfield2009}, the coupling parameter is expressed as the effective length $L_\mathrm{eff}^{-1}=\tfrac{1}{\omega_\mathrm{c}}\tfrac{d\omega_\mathrm{c}}{dx}$ of a resonator for which $|G|=\omega_\mathrm{c}/L_\mathrm{eff}$.
In this approach, the coupling parameter may be defined individually for each mechanical mode.

In all cases, the only  physically significant relation is the induced resonance frequency shift $\Delta \omega_\mathrm{c}$ upon a displacement $\mvec u(\mvec r)$, which in a perturbation theory treatment, can be calculated according to%
\footnote{For open systems, the integration boundaries should be chosen close to the surface of the resonator, to avoid divergence of the integral in the denominator. Alternatively, radiating fields escaping from the volume of interest have to be taken into account by a surface integral \cite{Lai1990}.}
\begin{equation}
\frac{\Delta\omega_\mathrm{c}}{\omega_\mathrm{c}}\approx
  \frac{1}{2}
  \frac
  {
{\int} {  \mathlarger| \mvec E(\mvec r)  \mathlarger|^2 \cdot
\left( \epsilon\left(\mvec r+\mvec u(\mvec r)\right) -\epsilon\left(\mvec r\right) \right) \, \mathrm{d}r^3 } 
  }
  {
{\int}{\mathlarger | \mvec E(\mvec r) \mathlarger |^2 \cdot \epsilon(\mvec r) \, \mathrm{d}r^3}
  }.
%\frac{\int|\vec{E}(\vec{r})|^{2}%\,\vec{\nabla}\varepsilon(\vec{r})\cdot\vec{u}(\vec{r})d^{3}r}{2\int|\vec{E}(\vec{r})|^{2}\,\varepsilon(\vec{r})d^{3}r},
\label{e:john}
\end{equation}
This expression is uniquely determined by the distribution of the dielectric constant $\epsilon$ and the (unperturbed) electric field $\mvec E$ and can, in most cases, be linearized in the small displacements $\mvec u$ and reduced to a surface integral over the relevant resonator boundaries \cite{Gillespie1995,Johnson2002}. 

In contrast to the unambiguous relation (\ref{e:john}), the equivalent displacement $x$ which is a-priori not a well-defined quantity, may be defined in an arbitrary way as discussed above.
An important consequence of this ambiguity in the choice of $x$ is the necessity to introduce an effective mass $\meff$ of the mechanical mode  \cite{Gillespie1995, Pinard1999} which retains the correct relation for the energy $U$ stored in the mechanical mode given a displacement~$x$,
\begin{equation}
  U=\frac{1}{2} \meff \Om^2 x^2.\label{energy}
\end{equation}
Evidently, rescaling $x\rightarrow \alpha x$ implies $\dwdx\rightarrow \dwdx/\alpha$ and $\meff\rightarrow \meff/\alpha^2$, so that the frequently quoted figures of merit $\dwdx$ and $\meff$ are individually only of very limited use. 
It is possible to eliminate this ambiguity by referring to the \emph{vacuum optomechanical coupling rate} 
\begin{equation}
 \vcr\equiv \dwdx \cdot \xzpf
\end{equation}
using terminology analogous to cavity quantum electrodynamics \cite{Kimble1994}. 
As expected from the Hamiltonian description (\ref{e:hint}), all optomechanical dynamics can be fully described in terms of this parameter.
As a consequence, parameters such as the cooling rate or optical spring effect can be derived upon knowledge of $\vcr$. 
For instance, the cooling rate in the resolved sideband regime is given by $\Gamma = 4 \langle \hat n_\mathrm{o} \rangle \vcr^2 / \kappa$, while the power to reach the SQL scales with $\kappa^2/\vcr^2$ with  the cavity loss rate $\kappa$.
Furthermore, the regime where $\vcr\sqrt{\langle \hat n_\mathrm{o} \rangle} \gg \kappa, \Gm$ ($\Gm$ is the mechanical damping rate) denotes the strong optomechanical coupling regime, in which parametric normal mode splitting occurs \cite{Marquardt2007,Dobrindt2008,Groblacher2009a}.

\section{Determination of the vacuum optomechanical coupling rate}

\begin{figure}[bth]
\begin{center}
{\includegraphics[width= \linewidth]{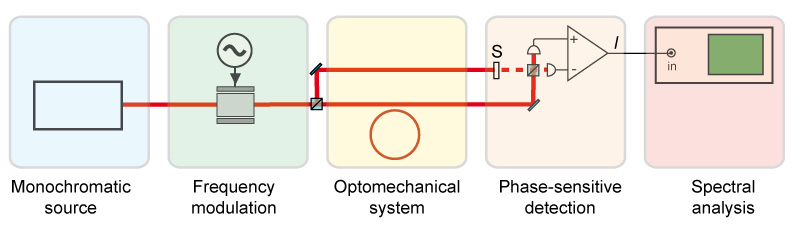}  }
\end{center}
\caption{Generic scheme for the determination of the vacuum optomechanical coupling rate.
The cavity optomechanical system is probed with a monochromatic source of electromagnetic waves which is frequency-modulated by a known amount at a Fourier frequency close to the mechanical resonance frequency. A phase-sensitive detector generates a signal $I$, which is analyzed with a spectrum analyzer. As two possible examples, a homodyne detector measures the phase difference between the arms in an interferometer, one of which accommodates the optomechanical system (shutter S open). If the shutter S is closed, the transmission signal $I$ of an optomechanical system is amplitude-dependent when the driving wave is detuned with respect to the optical resonance.  
}%
\label{f:detection}
\end{figure}

Our intention here is to detail and verify an experimental technique to determine $\vcr$ in a simple and straightforward manner for an arbitrary optomechanical system. 
The technique we present relies on the fluctuation-dissipation theorem, from which the double-sided (symmetrized) spectral density of displacement fluctuations of a mechanical oscillator in contact with a thermal bath at temperature $T$ can be derived 
\cite{Callen1951,Saulson1990,Gillespie1995},
\begin{align}
S_{xx}(\Og)=
\frac{\Gm\hbar\Og}{\meff}\frac{\coth\left(\frac{\hbar\Og}{2k_\mathrm{B} T}\right)}{(\Og^2-\Om^2)^2+ \Gm^2\Og^2}\approx
\frac{1}{\meff}\frac{2  \Gm k_\mathrm{B} T }{(\Og^2-\Om^2)^2+ \Gm^2\Og^2},
\end{align}
where $\meff$ is the effective mass corresponding to the choice of the normalization of $x$, as derived from (\ref{energy}), and the approximation is valid for large mechanical occupation numbers  $\langle n_\mathrm{m} \rangle \approx k_\mathrm{B} T/\hbar \Om\gg1$.
The displacement fluctuations are not directly experimentally accessible, but cause fluctuations of the cavity resonance frequency~$\omega_\mathrm{c}$ according to
\begin{align}
S_{\omega\omega}(\Og)=
\dwdx^2 S_{xx}(\Og)\approx
%\frac{\dwdx^2}{\meff}\frac{2  \Gamma_\mathrm{m}k_\mathrm{B} T }{(\Og^2-\Om^2)^2+ \Gamma_\mathrm{m}^2\Og^2}=
\vcr^2\cdot\frac{2\Om}{\hbar}\frac{2  \Gm k_\mathrm{B} T }{(\Og^2-\Om^2)^2+ \Gm^2\Og^2}.
\end{align}
If such a spectrum can be measured, it is straightforward to integrate it over Fourier frequencies resulting in the interesting relation
\begin{align}
 \langle  \delta \omega_\mathrm{c}^2\rangle
 =
 \int_{-\infty}^{+\infty}
S_{\omega\omega}(\Og) \frac{\mathrm{d}\Og}{2\pi}
=S_{\omega\omega}(\Om)\frac{\Gm}{2}=2\langle n_\mathrm{m} \rangle \vcr^2,
\label{e:main}
\end{align}
which implies that the knowledge of the mechanical occupation number and the cavity resonance frequency fluctuation spectrum are sufficient to directly derive $\vcr$.
The former is usually a straightforward task (provided dynamical backaction \cite{Braginsky2001,Braginsky2002} is ruled out, for example, by low probing powers), whereas the latter has to be implemented using a phase-sensitive measurement scheme as shown in figure \ref{f:detection}, since the required changes of the optical resonance frequency change the phase of the internal and hence output field.

Various methods for such measurements are available, such as direct detection when the probing laser is detuned, frequency modulation \cite{Drever1983}, polarization \cite{Hansch1980}, or homodyne spectroscopy \cite{Hadjar1999}.
In general, cavity resonance frequency fluctuations are transduced into the signal in a frequency-dependent manner, such that the spectrum of the measured signal (e.g. a photocurrent) $I$ is given by   
\begin{equation}
 S_{II}(\Og)=K(\Og) S_{\psi\psi}(\Og)=\frac{K(\Og)}{\Og^2}S_{\omega\omega}(\Og),
  \label{e:Sii}
\end{equation}
where $\psi$ is the phase of the intracavity field, which gets modulated upon a change of the resonance frequency.
Importantly, to exploit the relation (\ref{e:main}), it is necessary to convert the spectrum  $S_{II}(\Og)$ back into a frequency noise spectrum, so the function $K(\Og)\equiv K(\Og,\Delta,\etac,\kappa,P)$ (where $\Delta=\omega_\mathrm{l}-\omega_\mathrm{c}$ is the detuning from resonance, and $\etac=\kappa_\mathrm{ex}/\kappa$ is the coupling parameter quantifying the ratio of the cavity loss rate due to input coupling $\kappa_\mathrm{ex}$ compared to the total loss rate $\kappa$) has to be calibrated in \emph{absolute} terms.  
Experimentally, this can be accomplished by frequency-modulating the source that is used for measuring $S_{II}$
\cite{Hadjar1999,Gorodetsky2004,Schliesser2008}.
Under correct experimental conditions (cf.\ appendix) the phase modulation of the input field undergoes the exactly same transduction, so that
\begin{equation}
 S_{II}(\Og)=K(\Og) S_{\phi\phi}(\Og),
\end{equation}
where $S_{\phi\phi}(\Og)$ is the spectrum of input field phase modulation.
For the typical case of monochromatic input field phase modulation ($E_\mathrm{l}\propto e^{-i \phi_0 \cos\Omod t}$), the equivalent spectral density is given by 
$S_{\phi\phi}(\Og)= 2\pi\frac{1}{2}\left(\delta(\Og-\Omod)+\delta(\Og+\Omod)\right)\phi^2_0/2$.   

A laboratory spectrum analyzer will, of course, not resolve this spectrum but convolute it with its bandwidth filter lineshape $F(\Og)$, for example a Gaussian centered at $\Og=0$.
If a time-domain trace is analyzed using a discrete Fourier transform, the filter function would be given by the squared modulus of the Fourier transform of the time-domain windowing function.
We then have%
\begin{equation}
  S_{II}^\mathrm{meas}(\Og)=2\cdot F(\Og)\ast \left(K(\Og)\left(S_{\psi\psi}(\Og)+S_{\phi\phi}(\Og)\right)\right),
\end{equation}
where a factor of $2$ was introduced to account for the fact that most analyzers conventionally display single-sided spectra (this factor is irrelevant for our final result, however). 
The filter function is usually normalized and can be characterized by the effective noise bandwidth (ENBW), for which
\begin{equation}
 F(0)\cdot\mathrm{ENBW} =\int_{-\infty}^{+\infty} F(\Og) \frac{\mathrm{d}\Og}{2\pi}=1,
 \end{equation}
essentially absorbing differences in hardware or software implementations of spectral analysis \cite{Heinzel2002}.
Note, however, that this description is still idealized. In reality, correction factors may apply depending on details of the signal processing (such as averaging of logarithmic noise power readings) in the respective analyzer used, see e.g. \cite{Agilent1303}.

If the effective noise bandwidth is chosen much narrower than all spectral features in $S_{\psi\psi}(\Og)$, the measured spectrum is approximately
\begin{equation}
  S_{II}^\mathrm{meas}(\Og)
  \approx 2\cdot
  K(\Og) S_{\psi\psi}(\Og)+ \frac{\phi_0^2}{2} K(\Omod)
  \left(F(\Og-\Omod)+F(\Og+\Omod)\right).
  \label{e:combinedspectrum}
\end{equation}
Assuming that the signal at $\Omod$ is dominated by laser modulation, the transduction function $K$ can then be calibrated in absolute terms at this frequency, as in the relation
\begin{equation}
 { S_{II}^\mathrm{meas}(\pm\Omod)}
  \approx 
  \frac{ \phi_0^2}{2} K(\Omod)
  F(0)
  =
  \frac{ \phi_0^2}{2} \frac{K(\Omod)}{\mathrm{ENBW}}.  
  \label{e:calibpeak}
\end{equation}
all parameters except $K(\Omod)$ are known.
The spectral shape of $K$ is then needed to derive the spectrum $S_{\psi\psi}$ of interest.
This task can be accomplished either by repeating the measurement for all modulation frequencies of interest, or by theoretical
analysis, relating it to the conditions under which the measurements are taken.

As the simplest example, we may consider the case of direct detection, where the signal is the power $P$ ``transmitted'' through the optomechanical system. 
For a given input power $P_\mathrm{in}$, we find (see appendix) the transduction function:  
\begin{align}
	K_D(\Og) =\frac{S_{PP}}{P^2_\mathrm{in} S_{\psi\psi}}=\frac{S_{PP}}{P^2_\mathrm{in} S_{\phi\phi}}=
\frac{4\etac^2\kappa^2\Delta^2\Og^2(\Og^2+\kappa^2(1-\etac)^2)}{((\Og+\Delta)^2+\kappa^2/4)((\Og-\Delta)^2+\kappa^2/4)(\Delta^2+\kappa^2/4)^2}.
\end{align}
We assumed in this expression sufficiently low pump powers such that effects related to optomechanically induced tranparency (OMIT) \cite{Weis2010} are negligible.
For the more advanced homodyne spectroscopy we calculate (see appendix),  
\begin{align}
%\phi_\mathrm{LO}&=-\arctan\left(\frac{\eta_c\kappa\Delta}{\Delta^2+(1-2\eta_c)\kappa^2/4}\right)\nonumber\\
K_H(\Og) &= \frac{S_{HH}}{P_\mathrm{in}P_\mathrm{LO}S_{\psi\psi}}=\frac{S_{HH}}{P_\mathrm{in}P_\mathrm{LO}S_{\phi\phi}}=\nonumber\\
&= \frac{4 \kappa^2\etac^2 \Og^2(\Og^2(1-2\etac)^2\kappa^2/4 +(\Delta^2-(1-2\etac)\kappa^2/4)^2)}{((\Delta-\Og)^2+\kappa^2/4)((\Delta+\Og)^2+\kappa^2/4)(\Delta^2+(1-2\etac)^2\kappa^2/4)(\Delta^2+\kappa^2/4)}.
\end{align}
Advantageously, this method can also be applied with the laser tuned to resonance ($\Delta=0$), ruling out effects of dynamical backaction  \cite{Braginsky2001,Braginsky2002} even for larger laser powers.
In this special case 
\begin{align}
K_H(\Og)&=\frac{16 \etac^2 \Og^2}{\Og^2+\kappa^2/4}.
\end{align}
Another interesting case is $\Delta=\pm \Om$ in the resolved sideband regime \cite{Schliesser2008} with $\Om\gg \kappa$, where
\begin{align}
K_H(\Om)&\simeq 4\etac^2\left(1+\frac{\kappa^2(16\etac-13)}{16\Om^2} \right).
\end{align}

We would finally like to emphasize that in many experimental situations, the transduction function $K(\Og)$ is sufficiently constant over the range of frequencies, in which the mechanical response in (\ref{e:main}) is significant (several mechanical linewidths around $\Om$).
In this case, it can be simply approximated to be flat, $K(\Og)\approx K(\Om) \approx K(\Omod)$, and we get from equations
(\ref{e:main}), (\ref{e:Sii}), (\ref{e:combinedspectrum}) and (\ref{e:calibpeak})
\begin{align}
  \vcr^2
  	%&=\frac{1}{2 n} \langle \delta\omega^2 \rangle=\\
  	%&=\frac{1}{2 n}\int_{-\infty}^{+\infty} \frac{\Omega^2 }{K(\Omega)} S_{II}^\mathrm{meas}(\Omega)\frac{\mathrm{d}\Omega}{2 \pi}
	%\approx\frac{1}{2 n} \frac{\Omega_\mathrm{mod}^2 }{K(\Omega_\mathrm{mod})} \int_{-\infty}^{+\infty}S_{II}^\mathrm{meas}(\Omega)\frac{\mathrm{d}\Omega}{2 \pi}\\
	&\approx\frac{1}{2 \langle n_\mathrm{m}\rangle}\frac{\phi_0^2 \Omega_\mathrm{mod}^2 }{2}
		\frac
			{S_{II}^\mathrm{meas}(\Omega_\mathrm{m})\cdot \Gamma_\mathrm{m}/4}
			{S_{II}^\mathrm{meas}(\Omega_\mathrm{mod})\cdot \mathrm{ENBW}}.
				\label{e:simple}
\end{align}
The last fraction in expression (\ref{e:simple}) can be interpreted as the ratio of the integrals of the displacement thermal noise and the calibration peak as broadened by the filter function.
Note also that according to its definition the ENBW is given in direct frequency and not angular frequency as is $\Gamma_\mathrm{m}$.
 
\section{Application of the method}

\begin{figure}[hbt]
\begin{center}
{\includegraphics[width= \linewidth]{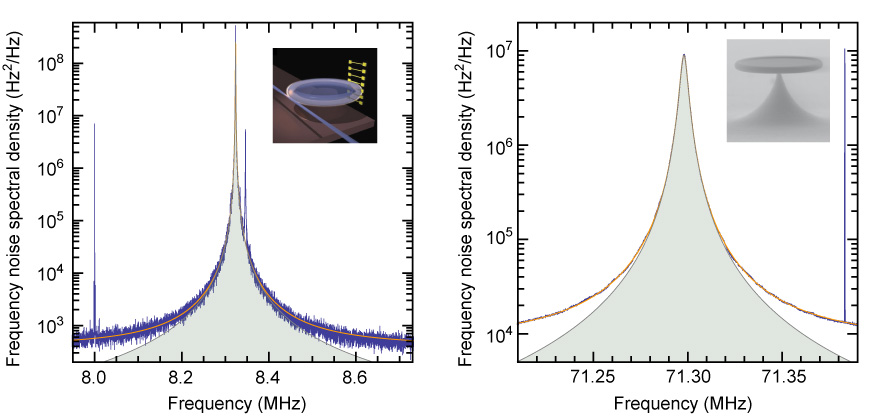}  }
\end{center}
\caption{
Measured (single-sided) frequency noise spectra $S_{\nu\nu}^\mathrm{meas}(\Og)$ (blue lines, $\nu\equiv \omega_\mathrm{c}/2\pi$) induced by a nanomechanical beam oscillator (8.3 MHz) coupled to a silica microtoroid \cite{Anetsberger2010} (left) and a radial breathing mode of a silica microtoroid (right), calibrated using laser frequency modulation at frequencies of 8 and $71.38\unit{MHz}$, respectively. Orange lines are Lorentzian fits, the gray area underneath them yield the optomechanical coupling rates multiplied with the occupation number of the oscillator. 
}
\label{f:spectra}
\end{figure}

As two examples, we have applied the discussed method to two different optomechanical systems shown in figure \ref{f:spectra}.
We first  consider the case of a doubly-clamped strained silicon nitride beam ($30\times 0.7\times 0.1 \unit{\mu m^3}$) placed in the near-field of a silica toroidal resonator \cite{Anetsberger2009}. 
This scheme allows sensitive transduction of the nanomechanical motion with an imprecision below the level of the standard quantum limit \cite{Anetsberger2010}.

Note that the coupling rate in such a configuration depends on the strength of the evanescent field at the location of the nanomechanical oscillator, which increases exponentially if the oscillator is approached to the toroid.
The data in figure \ref{f:spectra} show the single-sided cavity frequency noise spectral density 
$S_{\nu\nu}^\mathrm{meas}(\Og)\equiv S_{II}^\mathrm{meas}(\Og)\cdot(\Og/2\pi)^2/K(\Og)$ recorded using the homodyne technique.
The spectrum was calibrated by laser frequency modulation at the Fourier frequency of $8\unit{MHz}$.
The spectrum needs not be processed further, since the transduction function for homodyne detection with $\Delta=0$ is suffciently flat around the mechanical peak, $\Gm\ll\kappa$.
However, the cavity frequency noise spectrum as measured here contains also contributions from other noise mechanisms present in the silica toroidal oscillator, in particular  thermorefractive noise \cite{Gorodetsky2004} known to dominate frequency noise at lower ($\lesssim 10\unit{MHz}$) Fourier frequencies.
The integration of the spectrum is therefore exclusively performed on the Lorentzian corresponding to the mechanical mode (gray area), which can be easily isolated by fitting the noise spectrum with a model accommodating all known noise mechanisms including thermorefractive noise \cite{Schliesser2008b,Anetsberger2010}.
The result of this evaluation is $\langle \delta \omega_\mathrm{c}^2\rangle=\left(2 \pi\cdot 530\unit{kHz}\right)^2$, and the corresponding vacuum coupling rate is given by $\vcr=2 \pi\cdot  420 \unit{Hz}$.

Silica WGM microresonators such as spheres or toroids actually support structure-inherent micromechanical modes themselves \cite{Ma2007, Schliesser2008b}.
Figure \ref{f:spectra} also shows an equivalent cavity frequency noise spectrum measured on such a cavity using direct detection.
Again we neglect the frequency-dependence of $K$ due to the very narrow mechanical peak ($\Gm\ll \kappa, \Om$).
For the evaluation of $\vcr$, we remove backgrounds due to thermal noise from other modes, thermorefractive fluctuations or shot noise, by fitting a Lorentzian model with a flat background term.
Only the peak amplitude of the Lorentzian is then used with the relation (\ref{e:simple}).
A vacuum coupling rate of $\vcr=2\pi \cdot 570\unit{Hz}$ is obtained, which is interestingly very similar to the vacuum coupling rate of the nanomechanical beam.

\section{Conclusion}

In summary we have presented a simple method which allows direct measurements of the vacuum coupling rate $\vcr$ in optomechanical systems using frequency-modulation.
We have applied the method to two different devices, namely  doubly-clamped strained silicon nitride beam ($\vcr/2\pi\approx 10^3\unit{Hz}$) and the radial breathing mode in a silica toroidal resonator ($\vcr/2\pi\approx 10^3\unit{Hz}$).
It is interesting to note how widely this coupling rate can vary in different optomechanical systems, for example, we estimate  $\vcr/2\pi\sim 10^{-2}\unit{Hz}$ for (present) microwave systems \cite{Regal2008}, $\vcr/2\pi\sim 10^{0}\unit{Hz}$ in lever-based optical systems \cite{Arcizet2008a,Groblacher2009a},  
% $\vcr/2\pi\sim 10^{3}\unit{Hz}$ in silica whispering-gallery-mode microresonators  \cite{Ma2007, Schliesser2008b} 
and up to $\vcr/2\pi\sim 10^{5}\unit{Hz}$ in integrated optomechanical crystals \cite{Eichenfield2009}.
Together with the optical and mechanical dissipation rates and resonance frequencies (which do also vary widely), the coupling rate provides a convenient way to assess the potential of optomechanical platforms for the different phenomena and experiments within the setting of cavity optomechanics.

\section*{Acknowledgements:} M.~G. acknowledges support from the Dynasty foundation. This work was funded by an ERC\ Starting Grant SiMP, the EU (Minos) and by the DARPA/MTO ORCHID program through a grant from AFOSR and the NCCR of Quantum Photonics.

\begin{appendix}

\section{Transduction coefficients}

In this Appendix we show by direct calculations that the amplitude transduction coefficients of the input phase modulation and 
of the mechanical motion are  the same both in the case of either phase homodyne or direct amplitude detection.

The canonical equations describing dynamical behavior of the optomechanical system are the following:
\begin{align}
	\dot a = (i\Delta - i\dwdx x(t)- \kappa/2)a(t)+\sqrt{\etac\kappa}s_\mathrm{in}(t), \\
  \ddot x(t)+\Gm\dot x(t)+\Om^2 x(t) = -\hbar \dwdx|\bar a|^2, \nonumber
\end{align}
where $\bar a$ is the internal field in the cavity normalized so that $|\bar a|^2$ is the number of photons in the mode, $s_\mathrm{in}$ is the
input field related to the input power as $P_\mathrm{in}={\hbar\omega_\mathrm{o}}|s_\mathrm{in}|^2$, $\Delta$ is the detuning of the pump from optical resonance, 
$\kappa$ and $\Gm$ are total optical and mechanical damping rates, $\etac$ is the coupling strength equal to the ratio of coupling and 
total optical losses and $\Om$ is mechanical resonance frequency. We neglect in the calculations below the effects of dynamical and quantum backaction of 
the optical probe field on the mechanical subsystem, implying sufficiently low input power.    

Assume that the mechanical degree of freedom of the cavity harmonically oscillates  so that 
\begin{align}
	x(t)= x_0 \cos(\Om t) \, .
\end{align}

These oscillations will produce a periodic change of the optical eigenfrequency and hence the phase of the internal field with  $\psi_0 = x_0\dwdx/\Om$. If these oscillations are small ($|\psi_0| \ll 1$), the spectral components of the internal fields in the cavity and output field in the coupler will be:
\begin{align}
  a_x &= s_\mathrm{in}\sqrt{\etac\kappa}{\cal L}(0)\left(1-\frac{i\psi_0\Om {\cal L}(+\Om)}{2}e^{-i\Om t}-\frac{i\psi_0\Om {\cal L}(-\Om)}{2}e^{+i\Om t} \right), \\
	s_{x,out}&=s_\mathrm{in}-\sqrt{\etac\kappa} a_x, \nonumber\\
  {\cal L}(\Og)&= \tfrac{1}{-i(\Delta+\Og)+\kappa/2}.\nonumber
\end{align}
In the same way if the input field is weakly phase modulated ($s_\mathrm{in}\propto e^{-i \phi_0 \cos(\Omod t)}$) to obtain a calibration signal:
\begin{align}
  s_{\phi,\mathrm{in}} &= s_\mathrm{in}\left(1-\frac{i\phi_0}{2}e^{-i\Omod t}-\frac{i\phi_0}{2}e^{+i\Omod t} \right), \\
	a_{\phi}&=s_\mathrm{in}\sqrt{\etac\kappa}\left({\cal L}(0)-\frac{i\phi_0 {\cal L}(+\Omod)}{2}e^{-i\Omod t}-\frac{i\phi_0 {\cal L}(-\Omod)}{2}e^{+i\Omod t} \right).\nonumber\\
	s_{\phi,out}&=s_{\phi,\mathrm{in}}-\sqrt{\etac\kappa} a_\phi.
\end{align}
Dynamically these two modulations are not equivalent, but surprisingly produce the same transduction coefficient in terms
of spectral densities.

We discuss here two different methods to probe the output field, either by directly looking at optical power modulation on 
a detector with appropriate detuning $ \Delta$ or by measuring phase oscillations using homodyne detection at zero detuning, 
with the local oscillator signal obtained by splitting the input power. 
From these relations it is possible to calculate the spectral densities of the power detected directly at the output of the detector, and the homodyne signal, once the spectral densities of mechanical ($S_{xx}(\Omega)$) or laser phase ($S_{\phi\phi}(\Og)$) are known.
%Below we calculate the transduction coefficients between mechanical displacement noise on the one hand and homodyne power spectral densities from the other hand by formally replacing, correspondingly, $x_0^2\to S_{xx}(\Og)=\Om^2 S_{\psi\psi}(\Og)/\dwdx^2$, $\phi^2_0 \to S_{\phi\phi}(\Og)$, $|H(\Om)|^2 \to S_{HH}(\Og)$ and $|P_\mathrm{out}(\Om)|^2 \to S_{PP}(\Og)$.

\subsection{Direct detection}

Directly calculating the modulation amplitude of the output optical power on a detector $P_\mathrm{out}(\Og)=\hbar\omega_0|s_\mathrm{out}|^2_\Og$ (omitting detector efficiency to convert to a photocurrent)
we  find that
\begin{align}
	K_D(\Og)
		&=\frac{S_{PP}(\Og)}{P^2_\mathrm{in} S_{\psi\psi}(\Og)}=\frac{S_{PP}(\Og)}{P^2_\mathrm{in} S_{\phi\phi}(\Og)}=\nonumber\\
		&=\frac{4\etac^2\kappa^2\Delta^2\Og^2(\Og^2+\kappa^2(1-\etac)^2)}
				{((\Og+\Delta)^2+\kappa^2/4)((\Og-\Delta)^2+\kappa^2/4)(\Delta^2+\kappa^2/4)^2}\, .
\end{align}

In resolved sideband regime when $\Og>\sqrt{2}\kappa$ the signal at frequency $\Og$ is maximized when the pump is either tuned
on the slope or on one of mechanical sidebands:
\begin{align}
\Delta=\pm \frac{1}{2}\sqrt{2\Og^2-\kappa^2-2\sqrt{\Og^4-2\kappa^2\Og^2}}\simeq \pm \frac{\kappa}{2}\left(1+\frac{\kappa^2}{4\Og^2}\right),\\
\Delta=\pm \frac{1}{2}\sqrt{2\Og^2-\kappa^2+2\sqrt{\Og^4-2\kappa^2\Og^2}}\simeq \pm \Og\left(1-\frac{3\kappa^2}{8\Og^2}\right),\nonumber
\end{align}
producing the same signal
\begin{align}
K_D(\Og)=4\etac^2\left(1+\frac{\kappa^2(1-\etac)^2)}{\Og^2}\right).
\end{align}
When $\Og<\sqrt{2}\kappa$, the output is maximized when 
\begin{align}
\Delta&=\pm \frac{1}{6}\sqrt{12\Og^2+3\kappa^2}, \\
K_D(\Og)&=\frac{27\Og^2\etac^2\kappa^2(\Og^2+\kappa^2(1-\etac)^2)}{(\Og^2+\kappa^2)^3}. \nonumber
\end{align}

\subsection{Balanced homodyne detection}

In this case, the signal on the detector is proportional to the field amplitudes $s_\mathrm{LO}$ and $s_\mathrm{in}$ falling on the output beam splitter,
\begin{align}
\frac{H}{\hbar\omega_0}&=|\frac{i}{\sqrt{2}}s_\mathrm{out}-\frac{1}{\sqrt{2}}s_\mathrm{LO}|^2-|\frac{i}{\sqrt{2}}s_\mathrm{LO}-\frac{1}{\sqrt{2}}s_\mathrm{out}|^2\nonumber\\
	 &=i (s_\mathrm{LO}s^*_\mathrm{out}-s^*_\mathrm{LO}s_\mathrm{out})=i(s_\mathrm{LO}s^*_\mathrm{in}-s^*_\mathrm{LO}s_\mathrm{in})+i\sqrt{\etac\kappa}(a s^*_\mathrm{LO}-a^*s_\mathrm{LO})\nonumber\\
  &=2|s_\mathrm{LO}||s_\mathrm{in}|\sin(\phi_\mathrm{LO}-\phi_\mathrm{in})+ i\sqrt{\etac\kappa}(a s^*_\mathrm{LO}-a^*s_\mathrm{LO})
\end{align}

It is important to note here that if the homodyne phase difference $\phi_\mathrm{LO}-\phi_\mathrm{in}=0$ then this method directly probes the phase of the internal field and hence the mechanical oscillations. This regime can be provided in case of $\Delta=0$. If the detuning is
not equal to zero, the phase difference in an experiment is usually locked to have zero d.c. voltage. The local oscillator field is
obtained by splitting the input power and is equal to:
\begin{align}
s_{x,\mathrm{LO}}&=\frac{s_\mathrm{LO}}{s_\mathrm{in}}s_{x,\mathrm{in}}e^{i\phi_\mathrm{LO}}=s_\mathrm{LO}e^{i\phi_\mathrm{LO}},\\
s_{\phi,\mathrm{LO}}&=\frac{s_\mathrm{LO}}{s_\mathrm{in}}s_{\phi,\mathrm{in}}e^{i\phi_\mathrm{LO}}=s_\mathrm{LO}e^{i\phi_\mathrm{LO}}\left(1-\frac{i\phi_0}{2}e^{-i\Omod t}-\frac{i\phi_0}{2}e^{+i\Omod t} \right), \nonumber\\
\end{align}
where $\phi_\mathrm{LO}$ is chosen to suppress any d.c. voltage,
\begin{equation}
\phi_\mathrm{LO}=-\arctan\left(\frac{\etac\kappa\Delta}{\Delta^2+(1-2\etac)\kappa^2/4}\right).
\end{equation}
In this case, again
\begin{equation}
\frac{S_{HH}}{P_\mathrm{in}P_\mathrm{LO}S_{\psi\psi}}=\frac{S_{HH}}{P_\mathrm{in}P_\mathrm{LO}S_{\phi\phi}}
\end{equation}
and
\begin{equation}
K_H(\Og)= \frac{4 \kappa^2\etac^2 \Og^2(\Og^2(1-2\etac)^2\kappa^2/4 +(\Delta^2-(1-2\etac)\kappa^2/4)^2)}{((\Delta-\Og)^2+\kappa^2/4)((\Delta+\Og)^2+\kappa^2/4)(\Delta^2+(1-2\etac)^2\kappa^2/4)(\Delta^2+\kappa^2/4)}.
\end{equation}

If $\Delta=0$ which appears the most favorable case for homodyne detection, then:
\begin{align}
K_H(0)&=\frac{16 \etac^2 \Og^2}{\Og^2+\kappa^2/4}.
\end{align}

Another interesting case is $\Delta=\pm \Om$ in case of resolved sideband regime. In this case
\begin{align}
K_H(\pm\Om)&\simeq 4\etac^2\left(1+\frac{\kappa^2(16\etac-13)}{16\Om^2} \right).
\end{align}

\end{appendix}

\end{document}